\documentclass[aps,twocolumn,nofootinbib,amsmath,amssymb,showpacs,superscriptaddress]{revtex4-1}

\usepackage{graphics}
\usepackage{setspace}
\usepackage{graphicx}
\usepackage{dsfont}
\usepackage{slashed}

\usepackage[    bookmarks,
                 bookmarksopen = true,
                 bookmarksnumbered = true,
                 breaklinks = true,
                 linktocpage,
                 colorlinks = true,
                 linkcolor = blue,
                 urlcolor  = blue,
                 citecolor = blue,
                 anchorcolor = green,
                 hyperindex = true,
                 hyperfigures]
		{hyperref}        

\newcommand{\dd}{\textmd{d}}
\newcommand{\be}{\begin{equation}}
\newcommand{\ee}{\end{equation}}

\newcommand{\Z}{\mathcal{Z}}

\newcommand{\expv}[1]{\left \langle #1 \right \rangle}
\newcommand{\tr}{\textmd{tr}\,}

\usepackage{tabularx}
\newcolumntype{C}{>{\centering\arraybackslash}X}
\newcolumntype{R}{>{\raggedleft\arraybackslash}X}
\newcolumntype{L}{>{\raggedright\arraybackslash}X}

\begin{document}

\title{New class of compact stars: Pion stars}

\author{B.~B.~Brandt}
\affiliation{Institute for Theoretical Physics, Goethe Universit\"at Frankfurt, D-60438 Frankfurt am Main, Germany}
\author{G.~Endr\H{o}di}
\affiliation{Institute for Theoretical Physics, Goethe Universit\"at Frankfurt, D-60438 Frankfurt am Main, Germany}
\author{E.~S.~Fraga}
\affiliation{Instituto de F\'\i sica, Universidade Federal do Rio de Janeiro,
Caixa Postal 68528, 21941-972, Rio de Janeiro, RJ, Brazil}
\author{M.~Hippert}
\affiliation{Instituto de F\'\i sica, Universidade Federal do Rio de Janeiro,
Caixa Postal 68528, 21941-972, Rio de Janeiro, RJ, Brazil}
\affiliation{Instituto  de  F\'\i sica,  Universidade  de  S\~ ao Paulo,  Rua  do  Mat\~ ao,
Caixa Postal 1371,  Butant\~ a,  CEP  05508-090,  S\~ ao  Paulo,  SP,  Brazil}
\author{J.~Schaffner-Bielich}
\affiliation{Institute for Theoretical Physics, Goethe Universit\"at Frankfurt, D-60438 Frankfurt am Main, Germany}
\author{S.~Schmalzbauer}
\affiliation{Institute for Theoretical Physics, Goethe Universit\"at Frankfurt, D-60438 Frankfurt am Main, Germany}

\date{\today}

\begin{abstract}
  We investigate the viability of a new type of compact star whose main 
  constituent is a Bose-Einstein
  condensate of charged pions. Several different setups are considered, where 
  a gas of charged leptons and neutrinos is also present. 
  The pionic equation of state is obtained
  from lattice QCD simulations in the presence of an isospin chemical potential,
  and requires no modeling of the nuclear force.
  The gravitationally bound configurations of these systems are found by solving the 
  Tolman-Oppenheimer-Volkoff equations. We discuss weak decays 
  within the pion condensed phase
  and elaborate on the generation mechanism of such objects. 
\end{abstract}

\maketitle

\section{Introduction}

Compact stellar objects offer deep insight into the physics of elementary particles in dense environments
through the imprint of merger events on the electromagnetic and gravitational 
wave spectra~\cite{TheLIGOScientific:2017qsa}. 
The theoretical description of compact star interiors requires full
knowledge of the equation of state (EoS) of nuclear matter and involves 
the non-perturbative solution of quantum chromodynamics (QCD), the theory
of strongly interacting quarks and gluons. 
However, first-principle methods (most notably, lattice QCD simulations) 
are not available for high baryon densities -- 
consequently, the EoS of neutron stars
necessarily relies on a modeling of the nuclear force. 
Here we propose a different scenario, where the neutron density vanishes and 
a Bose-Einstein condensate of charged pions (the lightest excitations in QCD)
plays the central role instead. 
This setting can be approached by first-principle methods and leads to a new
class of compact objects that we name pion stars. 
As we demonstrate, under certain circumstances 
pion star matter can indeed exhibit gravitationally bound configurations.

The most prominent representatives of compact stellar objects are 
neutron stars. 
The prediction of their existence~\cite{Landau:1932} and their association 
to the relics of core-collapse supernovae~\cite{Baade:1934} anticipated their
serendipitous discovery by more than three decades~\cite{Hewish:1968bj}. 
Today, more than 2600 pulsars,
rotation-powered neutron stars, are known and listed in the ATNF pulsar
database. 
However, the known pulsars are only the tip of the
iceberg, as approximately a billion neutron stars 
are likely to exist within our galaxy. 
Together with other compact objects, they can be exposed by signatures 
from their companion stars or by gravitational wave emission, revealing information
on their structure and composition.
While neutron star matter consists mostly of neutrons and protons (baryons) and, thus, 
features high baryon density, the proposed pion stars are substantially different.
Their strongly interacting component is 
characterized by zero baryon density and high isospin charge. Unlike neutron 
star matter, this system
is amenable to lattice QCD simulations using standard Monte-Carlo algorithms~\cite{Son:2000xc}, giving direct access to the EoS -- i.e., the 
relation between the pressure $p$ and the energy density $\epsilon$. 
  
Pion stars can be placed in the larger class of boson stars. Throughout their 
long history~\cite{Wheeler:1955zz,Kaup:1968zz,Jetzer:1991jr}, boson stars were
assumed, for example, to contain hypothetical elementary particles that would be either
free~\cite{Kaup:1968zz} or weakly interacting~\cite{Colpi:1986ye} scalars. 
Boson stars were also 
associated with Q-balls or Q-clouds -- non-topological solutions in scalar field theories~\cite{Coleman:1985ki,Alford:1987vs}.
Typically being much heavier and more extended than other compact objects, it was expected 
that boson stars might mimic black holes or serve as candidates for 
dark matter within galaxies~\cite{Liebling:2012fv}. 
Unlike boson stars considered previously, pion
stars have no need for any beyond Standard Model constituents. 
We also note that the presently discussed pion stars differ from neutron stars 
with a pion condensate core -- a setting which
has been explored in great detail in the past, see, e.g.\ Refs.~\cite{Migdal:1979je,pistar_1984,Migdal:1990vm}.

The Bose-Einstein condensation of charged pions involves the accumulation of isospin charge
at zero baryon density and zero strangeness. 
In QCD, isospin is conserved such that pion condensation can be triggered by 
an isospin chemical potential $\mu_I/2=\mu_u=-\mu_d$ that couples
to the third component of isospin and thus oppositely to the up 
and down quark flavors, and induces opposite quark densities $n_I=n_u=-n_d$. 
Such a difference in the light quark chemical 
potentials can arise in the early universe if a lepton asymmetry is present.
Indeed, in an electrically neutral system, an asymmetry between neutrino and antineutrino densities requires $\mu_I\neq0$~\cite{Schwarz:2009ii}. A sufficiently high 
lepton asymmetry can drive the system into the pion condensed phase~\cite{Abuki:2009hx} as the 
temperature $T$ drops.
Whether pion condensation takes place in the early universe depends on the 
initial conditions -- 
constrained by observations of the lepton asymmetry~\cite{Wygas:2018otj} --
and the subsequent evolution of 
the system in the QCD phase diagram in the $T-\mu_I$ plane. The structure of 
this phase diagram has been determined recently using lattice simulations~\cite{Brandt:2017oyy}.

\section{QCD sector}

As mentioned above, to describe pion condensation we can consider QCD with $\mu_I\neq0$, but
zero baryon and strangeness chemical potentials.
The low-energy effective theory of
this system is chiral perturbation theory ($\chi$PT), 
which operates with pionic degrees of freedom. According to $\chi$PT~\cite{Son:2000xc},
at zero temperature pions condense if $\mu_I\ge m_\pi$, where $m_\pi$ is the 
pion mass in the vacuum.\footnote{We note that here we follow a different 
convention compared Refs.~\cite{Endrodi:2014lja,Brandt:2017zck,Brandt:2017oyy}, 
where the threshold chemical potential equals $m_\pi/2$.}
Beyond this threshold the $\mathrm{U}(1)_{\tau_3}$ part 
of the chiral symmetry of the light quark action 
is broken spontaneously by the pion condensed ground state. 
The corresponding phase transition is 
of second order and manifests itself in a pronounced rise of the isospin 
density $n_I$ beyond the critical point~\cite{Son:2000xc}.  
The condensed phase exhibits nonzero energy density $\epsilon_\pi$ and, due to 
repulsive pionic interactions, nonzero pressure $p_\pi$. 
Besides isospin, the ground state also carries a non-vanishing electric charge 
density $n_Q = n_u\cdot q_u/e + n_d\cdot q_d/e = n_I$, 
where the fractional electric charges of the quarks $q_u=-2q_d=2e/3$ enter,
with $e>0$ being the elementary charge.\footnote{To relate the charge density to the isospin density, we assume
that the only charged states that contribute to the pressure have zero baryon number
and zero strangeness. 
This is indeed the case in the $T\to0$ limit if the isospin chemical 
potential is sufficiently small so that heavier charged hadrons are 
not excited. The strongest constraint is given by $\mu_I<m_K \approx 3.6 \,m_\pi $, 
where $m_K$ is the kaon mass, and is fulfilled in the following calculations.}
Without loss of generality we can assume $\mu_I>0$ so that the electric charge 
density is positive.

The isospin density $n_I$ and the pion condensate $\sigma_\pi=\langle\bar u \gamma_5 d - \bar d \gamma_5 u\rangle$ are obtained 
as expectation values involving the Euclidean
path integral over the gluon and quark fields discretized on a space-time lattice. 
The positivity of the measure in the path integral~\cite{Kogut:2002zg} ensures 
that standard importance sampling methods are applicable. 
Since the spontaneous symmetry breaking associated to pion condensation does not occur in a finite 
volume, the simulations are performed by introducing a pionic source parameter $\lambda$
that breaks the $\mathrm{U}(1)_{\tau_3}$ symmetry explicitly~\cite{Kogut:2002zg}.
Physical results are obtained 
by extrapolating this auxiliary parameter to zero. To facilitate a controlled 
extrapolation, we improve our observables using the approach discussed in Ref.~\cite{Brandt:2017oyy}.
The details of our lattice setup are described in Appendix~\ref{app:1}.

\begin{figure}[t]{
 \centering
 \vspace*{-.2cm}
 \includegraphics[width=8.5cm]{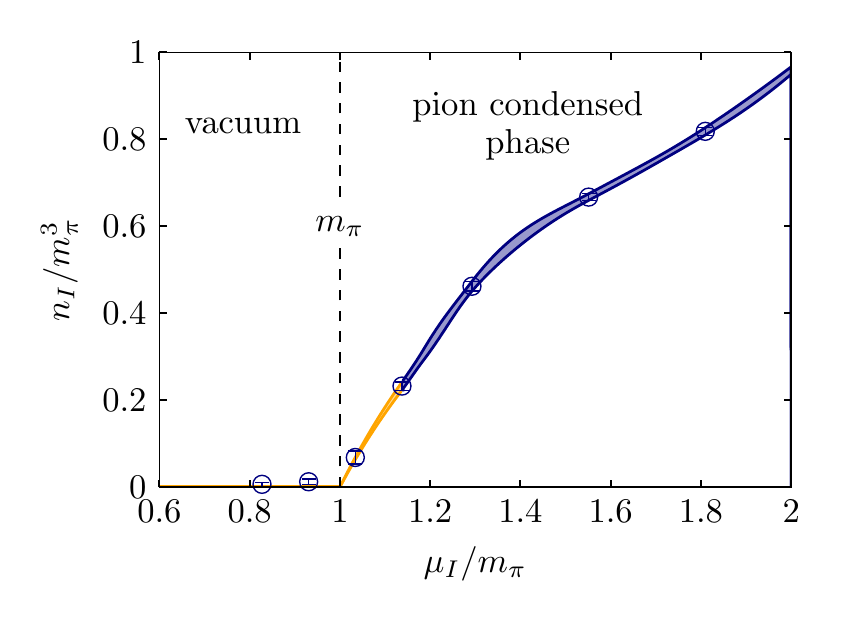}
 \vspace*{-.2cm}
 \caption{\label{fig:nI} Phase transition between the vacuum and the pion condensed phase, as exhibited by the isospin density. The lattice data (blue points) are fitted using
 $\chi$PT (yellow curve) and matched to a spline interpolation (blue curve).}
}
\end{figure}

The results of the $\lambda\to0$ extrapolation of the isospin density are shown in Fig.~\ref{fig:nI} as a function of the isospin chemical potential. 
The data clearly reflect the phase transition to the pion condensed 
phase at $\mu_I=m_\pi$. Due to effects from the finite volume and the small but
nonzero temperature employed in our simulations, the density just below $\mu_I=m_\pi$
is not exactly zero. To approach the thermodynamic and $T=0$ limits consistently, we 
employ $\chi$PT. In particular, we set the density to 
zero below $m_\pi$ and fit 
the lattice data to the form predicted by $\chi$PT around the critical chemical potential, 
see Appendix~\ref{app:2}. This involves fitting the pion decay constant, for which we obtain $f_\pi=133(4)\textmd{ MeV}$, in excellent agreement with its physical value.
Matching the fit to a spline 
interpolation of the lattice results at higher isospin chemical potentials
gives the continuous curve shown in Fig.~\ref{fig:nI}. 
Using standard thermodynamic relations (for details see Appendix~\ref{app:2}), 
the resulting $n_I(\mu_I)$ curve is used to calculate the EoS, 
shown in Fig.~\ref{fig:eos} below.

\section{Electroweak sector}
\label{sec:ew}

\begin{figure}[t]
 \centering
 \includegraphics[width=8.33cm]{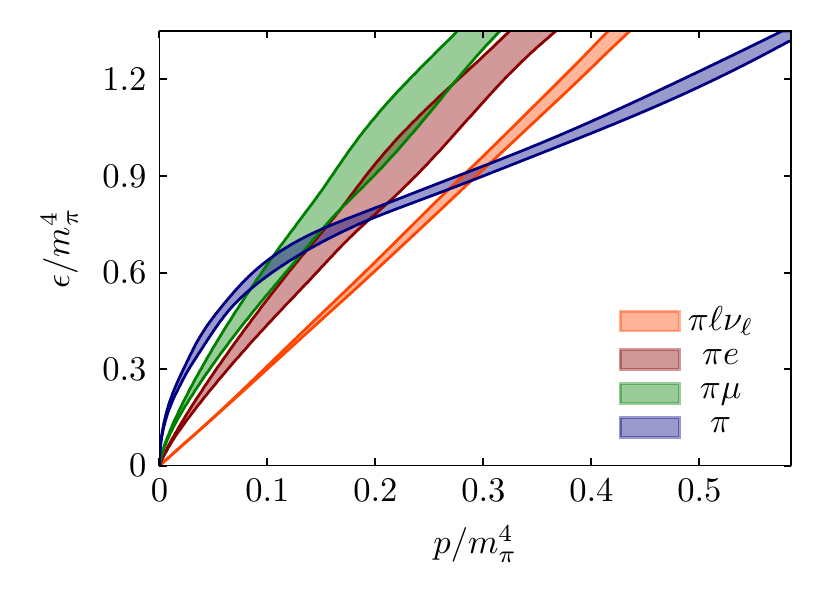}
 \caption{\label{fig:eos} Equation of state in the pion
 condensed phase in the QCD sector and for the electrically neutral 
 systems also including leptons (either muons or electrons) and neutrinos.
 The width of the curves 
 incorporates 
 statistical uncertainties as well as the uncertainty in the lattice pion mass for 
 the pion-lepton systems. }
\end{figure}

We consider the scenario where the pion condensate is neutralized by 
a gas of charged leptons with mass $m_\ell$. 
In the present approach we assume leptons to be free 
relativistic particles. A systematic improvement over this assumption is possible
by taking into account $\mathcal{O}(e^2)$ electromagnetic effects perturbatively, both in the electroweak sector and in lattice QCD simulations.\footnote{For charged leptons, this involves two-loop diagrams with an internal photon propagator, while in QCD a vacuum polarization diagram with two external photon legs is required.}
The lepton density $n_\ell$
is controlled by a lepton chemical potential $\mu_\ell$,
from which the leptonic contribution to the pressure $p_\ell$ and to the energy density 
$\epsilon_\ell$ can be obtained, similarly to the QCD sector. 
We require local charge neutrality to hold, $n_I + n_\ell = 0$, 
which uniquely determines the lepton chemical potential in terms of $\mu_I$.
The corresponding EoS for electrons ($\ell=e$) and for muons ($\ell=\mu$) 
is also included in Fig.~\ref{fig:eos}.
We mention that this setup was also investigated in Ref.~\cite{Andersen:2018nzq} and 
a similar construction, assuming a first-order phase transition for pions, 
was discussed in Ref.~\cite{Carignano:2016lxe}.

In the vacuum phase, charged pions decay weakly into leptons, with 
a characteristic lifetime of $\tau_{\rm vac}\approx10^{-8}\textmd{ s}$. 
In the condensed phase, the analogous weak process is quite different.
Since the spontaneously broken symmetry group corresponds to the 
local gauge group of electromagnetism, the pion condensed phase is 
a superconductor, where the Goldstone mode is a 
linear combination of the electric charge eigenstates $\pi^+$ and $\pi^-$~\cite{Son:2000xc}. 
In the presence of dynamical photons (in the unitary gauge) 
this mode disappears from the spectrum
via the Higgs mechanism~\cite{Migdal:1990vm}, at the cost of 
a nonzero photon mass $m_\gamma\propto e |\sigma_\pi|$. 
In addition, the other linear combination of $\pi^+$ and $\pi^-$ 
develops a mass above $\mu_I$~\cite{Son:2000xc} and is not excited if 
the temperature is sufficiently low. 
Thus, there is no light, electrically charged excitation that would decay weakly.
However, besides condensation in the pseudoscalar channel, the ground state also 
exhibits an axial vector condensate $\sigma_A=\langle\bar u \gamma_0\gamma_5 d + \bar d \gamma_0 \gamma_5 u\rangle/2$
that couples directly to the charged weak current as we discuss in Appendix~\ref{app:3}.

In Fig.~\ref{fig:sigma_a} we show first lattice results for $\sigma_A$. 
The measurements at different values of the auxiliary pion source parameter
are extrapolated to $\lambda\to0$ using an approach similar to that of Ref.~\cite{Brandt:2017oyy}, employing a generalized Banks-Casher relation
that we derive in Appendix~\ref{app:1}. 
Fig.~\ref{fig:sigma_a} also includes the $\chi$PT prediction~\cite{Brauner:2016lkh}, for which we use $f_\pi$
as obtained above for the fit of $n_I$. The results clearly show $\sigma_A>0$ in the 
condensed phase and a nice agreement between the two approaches.
The coupling of $\sigma_A$ to the weak current results in the depletion of the condensate
and the production 
of charged antileptons and neutrinos.
The characteristic lifetime $\tau$ of this process 
is calculated perturbatively in Appendix~\ref{app:3}. Normalized by the 
vacuum value $\tau_{\rm vac}$, we find that the lifetime takes the form
\be
\mu_I>m_\pi: \quad\frac{\tau}{\tau_{\rm vac}} = 
\frac{\mu_I^3}{m_\pi^3} \left[\frac{1-m_\ell^2/m_\pi^2}{1-m_\ell^2/\mu_I^2}\right]^2\,,
\label{eq:taures}
\ee
where the $\chi$PT prediction for $\sigma_A$ was used. 

Although suppressed deep in the condensed phase (as $\mu_I^{-3}$), 
weak decays therefore reduce the isospin charge of the 
system and create neutrinos $\nu_\ell$. For a high enough density of charged leptons and 
pions, the scattering cross section might be enhanced sufficiently 
to trap these neutrinos. Specifically, the conversion process $\nu_\ell\to \ell$ becomes possible 
where the neutrino couples to the condensate and transfers momentum to it. 
In addition, one also expects the cross section for $\nu_\ell \ell$ scattering 
to increase.

Thus, when the weak interactions are included, a consistent description of pion stars 
requires the inclusion of neutrinos.
Therefore we consider the scenario where a gas of 
neutrinos -- described by a density $n_{\nu_\ell}$ and a corresponding 
chemical potential $\mu_{\nu_\ell}$ -- is also present in the system. At weak equilibrium,
$\mu_{\nu_\ell}=\mu_I+\mu_\ell$,
this setup can maintain a pion condensate for high neutrino density, as was already 
shown in Ref.~\cite{Abuki:2009hx}. 
In this case, neutrinos also contribute to the pressure
and to the energy density by the amounts $p_{\nu_\ell}$ and $\epsilon_{\nu_\ell}$, respectively.
Since there are two leptonic pion decay channels, we consider both electrons and muons, as well as their respective neutrinos in our calculations.
Chemical equilibrium among the two families, resulting from 
neutrino oscillations, corresponds to $\mu_{\nu_\mu} = \mu_{\nu_e}$ or, equivalently,
$\mu_\mu = \mu_e$.
The EoS for this setup is also indicated in Fig.~\ref{fig:eos}.

\begin{figure}[t]{
 \centering
 \includegraphics[width=8.5cm]{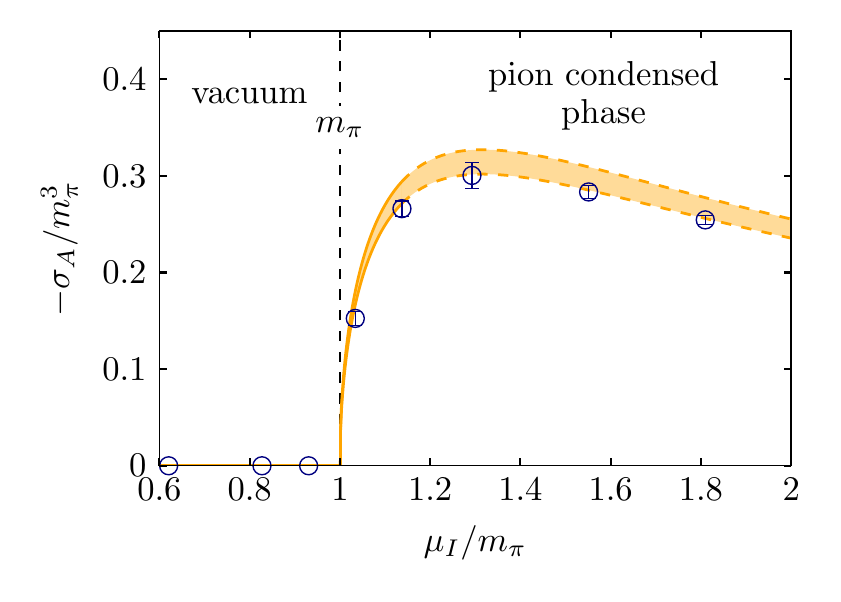}
 \caption{\label{fig:sigma_a} 
 Lattice data for the axial vector condensate, extrapolated to vanishing 
 pion source, $\lambda=0$, using our improvement program (blue points).
 The results are compared to the $\chi$PT prediction (yellow curve)~\cite{Brauner:2016lkh}.}
}
\end{figure}

\section{Gravity sector}

Using the resulting different 
equations of state, the mass $M$ and radius $R$ of pion stars can be computed by solving the
Tolman-Oppenheimer-Volkoff (TOV) equations~\cite{Tolman:1939jz,Oppenheimer:1939ne}, which 
describe hydrostatic equilibrium
in general relativity, assuming spherical symmetry. 
Our implementation is detailed in Appendix~\ref{app:2}. Further 
stability analyses are performed by requiring the star to be robust against 
density perturbations~\cite{glendenning2000compact} and radial oscillations.
The latter involves checking
whether unstable modes exist by solving the corresponding
Sturm-Liouville equation~\cite{Bardeen}. For more details on this analysis, 
see Appendix~\ref{app:2}.
Fig.~\ref{fig:mr1} shows the resulting mass-radius relations 
for pion stars of different compositions. 
The electrically charged pure pion stars\footnote{Our preliminary results for this case were presented in Ref.~\cite{Brandt:2017zck}.} have masses comparable to ordinary neutron stars,
but an order of magnitude larger radii. The inclusion of leptons (either electrons 
or muons) increases both the
masses and the radii considerably. Typically, the pion-electron configurations 
can be as heavy as intermediate-mass black holes~\cite{Pasham:2015tca}, 
whereas their radii are comparable to those of regular stars~\cite{Torres:2009js}.

\begin{figure}[t]
 \centering
 \includegraphics[height=6.2cm]{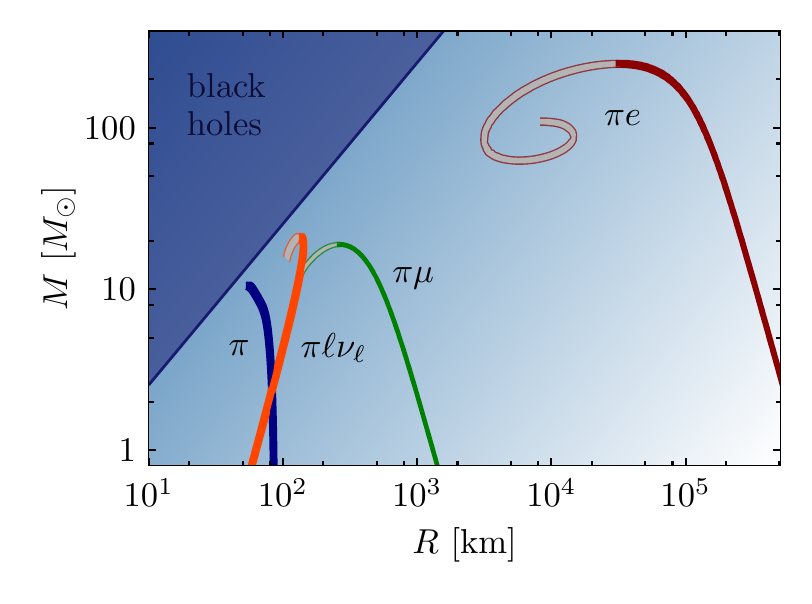}
 \vspace*{-.1cm}
 \caption{\label{fig:mr1} Mass-radius relations of various scenarios for pion stars.
 Shown is a pure pion star ($\pi$), a pion-electron ($\pi e$) and a pion-muon system ($\pi\mu$), together with a system containing both lepton families in chemical equilibrium ($\pi\ell\nu_\ell$).
 The filled (open) segments mark the 
 gravitationally stable (unstable) solutions (for details see the text). 
 The dark blue area marks 
 the region excluded by causality and the background color represents 
 the compactness $\beta\propto M/R$ of the objects (darker colors indicate more compact stars).
 The width of the curves indicates statistical errors
 and the uncertainty in the lattice pion mass.}
\end{figure}

In addition, we also considered a mixture of electrons and muons 
in chemical equilibrium by setting their respective chemical potentials equal. 
We observed that the gravitationally stable configurations for the latter setup 
cannot maintain a muonic component and are thus identical to those for 
the pion-electron system. Finally, the pion-lepton-neutrino scenario (with two 
lepton families in chemical equilibrium) again results in 
moderate masses and radii. We note that in this last case the star radius 
is defined by the point where the pressure of pions and of charged leptons vanishes,
while $p_{\nu_\ell}$ is still nonzero. 
These configurations may therefore be viewed as a pion-lepton star in a
neutrino cloud. 

Such a cloud, in the form of a background of
degenerate
neutrinos could be present for a high leptochemical potential in the
early universe,
a possible cosmological scenario for temperatures below the QCD transition,
as discussed in Ref.~\cite{Wygas:2018otj}.
Astrophysical neutrino clouds (with massive neutrinos) in the form of a Fermi star 
would be stable on galactic scales, see e.g.\ Ref.~\cite{Narain:2006kx}.
On the other hand, an unstable expanding neutrino cloud would lead to pion star
configurations which are
subject to evaporation near the border of the pion condensate. Consequently,
the escaping neutrinos will continuously be replaced by the ones resulting 
from the decay of the condensate in the outer layers. 
The details of such an evaporation process will predominantly depend on the 
(density-dependent) pion lifetime, the neutrino mean free path and the 
radius of the star. This calculation
is outside the scope of the present paper.

\begin{figure}[t]
\vspace*{-.2cm}
\hspace*{-.5cm}
\includegraphics[height=6.3cm]{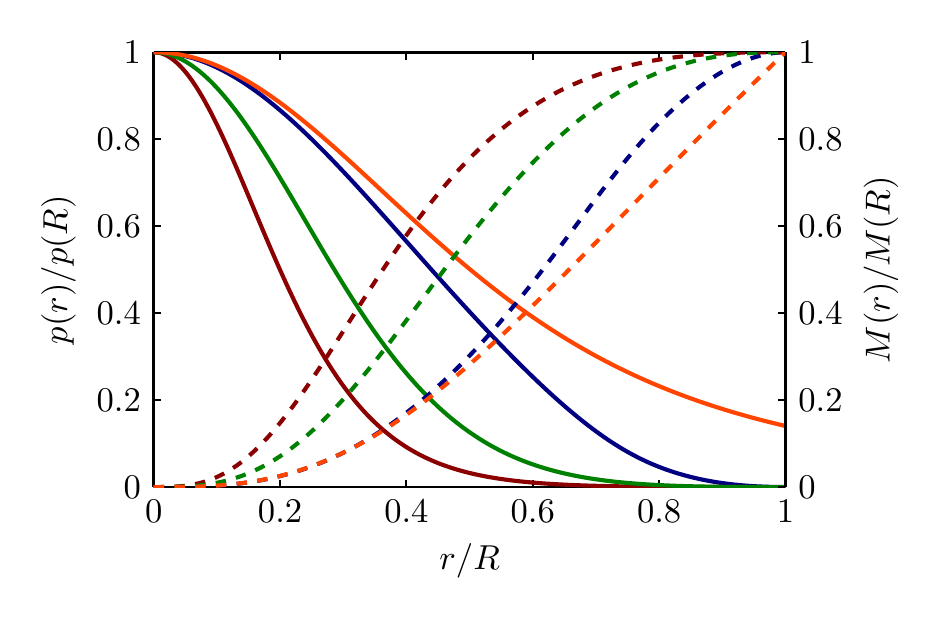}
\caption{Integrated mass (dashed) and total pressure (solid) within the $\pi$ (blue), $\pi e$ (red), $\pi \mu$ (green), $\pi\ell\nu_\ell$ (orange) stars at their respective maximum mass, c.f. Fig.~\ref{fig:mr1}. The mass accumulates to its maximum value, whereas the pressure drops towards the boundary of the star. In the presence of neutrinos, a finite pressure remains at the boundary of the condensate region.}
\label{fig:profile}
\end{figure}

\begin{table*}
\centering
\begin{minipage}{17.8cm}
\caption{Relevant properties of possible pion star compositions at their respective maximum mass. The last row correponds to the scenario with minimal neutrino pressure and $\mu_{\ell,c}/m_\ell$ is the relative electron chemical potential in this case.
}
\label{tab:star_properties_at_max}
\begin{tabularx}{\textwidth}{clCCCCCCc}
\hline\hline
\\[-0.3cm]
\hspace*{0.4cm}&composition& $M^{\rm max}\,[M_\odot]$ & $R^{\rm max}\,[\mathrm{km}]$ & $\epsilon_c^{\rm max}\,[\mathrm{MeV\!/fm}^3]$	& $p_c^{\rm max}\,[\mathrm{MeV\!/fm}^3]$	&$\mu_{I,c}^{\rm max}/m_\pi - 1$&$\mu_{\ell,c}^{\rm max}/m_\ell-1$&\hspace*{0cm}
\\[0.1cm]\hline
\\[-0.3cm]
&$\pi$	&10.5(5)	&55(3)	&57(5)	&25(3)	&1.068(4)	& - &\\
&$\pi e$	&250(10)	&$3.3(2)\times 10^{4}$	&$4.5(4)\times 10^{-5}$	&$3.5(4)\times 10^{-7}$	&$5.59(2)\times 10^{-7}$	& 7.4(2) &\\
&$\pi \mu$	&18.9(4)	&267(8)	&2.7(3)	&0.22(3)	&$1.623(5)\times 10^{-2}$	&0.58(3) &\\
&$\pi\ell\nu_\ell|_{\mu_\mu=\mu_e}$	&20.8(9)	&137(6)	&7.5(7)	&2.3(2)	&$4.13(2)\times 10^{-3}$	&160(4)&\\
&$\pi\ell\nu_\ell|_{\textmd{min.\ }p_{\nu_\ell}}$	&28(2)	&193(8)	&3.7(4)	&1.1(1)	&$3.77(2)\times 10^{-3}$	&155(4)&\\[.05cm]\hline\hline
\end{tabularx}
\end{minipage}
\end{table*}

We point out that Fig.~\ref{fig:mr1} reflects the $R \sim {\rm const.}$ behavior for 
pure pion stars (with masses below $7 \,M_\odot$) -- a telltale sign for
an interaction-dominated EoS. The slope changes by the addition of leptons, 
scaling as $M R^3 \sim {\rm const.}$, similarly to stars made of fermions.
Having a non-vanishing pressure at the boundary from neutrinos leads to a $M R^{-3} \sim {\rm const.}$ behavior for the $\pi\ell\nu_\ell$ configuration, reminiscent of 
a self-bound star with constant density.
Finally we also considered 
the $\pi\ell\nu_\ell$ system which minimizes 
the neutrino pressure at the star surface (and with it, the evaporation rate 
for the scenario of an expanding neutrino cloud). This condition was found to be met 
if $\mu_\mu=-m_\mu$ throughout the star. 
The corresponding results for $M(R)$ lie 
very close to the orange curve shown in Fig.~\ref{fig:mr1} with somewhat higher masses and 
radii. The profiles for the pressure and for the integrated energy density 
of the maximal mass configurations are plotted in Fig.~\ref{fig:profile}. 
Finally, 
an overview of the maximum masses and the corresponding radii is provided in Table.~\ref{tab:star_properties_at_max}, together with the central values for the 
energy density $\epsilon_c$, the pressure $p_c$ and the chemical potentials $\mu_{I,c}$ and 
$\mu_{\ell,c}$.

\section{Conclusions}

Pion stars provide a potential new class of compact objects, one that is made of a
Bose-Einstein condensate of charged pions and a gas of leptons, being significantly 
different from neutron stars and white dwarfs both in their structure and gross features. 
Pion condensation might have occured in the early universe if large lepton 
asymmetries were present~\cite{Schwarz:2009ii,Abuki:2009hx,Wygas:2018otj}, serving as a primordial production 
mechanism for pion stars. 
These new compact objects might be revealed 
by the characteristic neutrino and photon spectra stemming from their evaporation
or, if they survive sufficiently long, 
by signatures from companion stars via gravitational waves.

In the present paper we described the construction of pion stars and identified the key 
issues that concern their viability. There
are open questions regarding the lifetime of pion stars, 
related to the question of weak stability, the possibility of neutrino trapping and 
the evaporation processes at the surface. 
The present analysis can be improved by addressing these issues in more detail and 
also by generalizing the calculation to nonzero temperatures, thereby making 
the contact to the potential primordial production mechanism more direct. 
Keeping these issues in mind, pion stars provide the first example in which the mass and 
radius of a compact stellar object can be determined from first principles.
Furthermore, even if they happen to be short-lived, pion stars could constitute 
the first known case of a boson star and, remarkably, one with no need for any beyond 
Standard Model physics.\\

\noindent
{\bf Acknowledgments}
E.S.F.\ and M.H.\ are partially supported by CNPq, FAPERJ and INCT-FNA Proc.\ No.\ 464898/2014-5. M.H.\ was also supported by FAPESP Proc.\ No.\ 2018/07833-1 and Capes Procs.\ No.\ 
88887.185090/2018-00, via INCT-FNA, and 88881.133995/2016-01, which was determinant
for his participation in this work. 
B.B.B., G.E.\ and S.S.\ acknowledge support from the DFG 
(EN 1064/2-1 and SFB/TRR 55).
The authors are grateful to Jens O. Andersen and Misha Stephanov for inspiring discussions and useful comments.
The authors also thank Oliver Witham for a careful reading
of the manuscript, Alessandro Sciarra for help in creating the plots.\\

\appendix

\section{Lattice setup and improvement}
\label{app:1}

In the presence of the isospin chemical potential $\mu_I$ 
and the auxiliary pionic source parameter $\lambda$, 
the fermion matrices for the light and strange quarks read
\be
M_{ud} = \slashed{D}(\mu_I) + m_{ud} \mathds{1}
+ i\lambda \gamma_5\tau_2\,,\quad
M_s = \slashed{D}(0) + m_s\,,
\label{eq:Ml}
\ee
where $\slashed{D}$ is the Dirac operator,
\be
\slashed{D}(\mu_I) = \slashed{D}(0)\mathds{1}+ \frac{\mu_I}{2} \gamma_0 \tau_3 \,,
\ee
$\tau_a$ denote the Pauli matrices acting in flavor space and 
$m_{ud}$ and $m_s$ are the light and strange quark masses, respectively.
We discretize $\slashed{D}$ using the rooted staggered formulation, 
so that the Euclidean path integral over the gauge field $A_\mu$ becomes
\be
\Z = \int \mathcal{D}A_\mu \,[\det M_{ud}]^{1/4} \, [\det M_s]^{1/4} \,e^{-S_g}\,,
\label{eq:Z}
\ee
where $S_g$ is the gluonic part of the QCD action, for which we use the tree level-improved
Symanzik discretization. For the fermion matrices we employ stout smearing of the 
gauge fields. 
The determinants of $M_{ud}$ and of $M_s$ are positive~\cite{Son:2000xc,Kogut:2002zg}, allowing 
for a probabilistic interpretation and standard Monte-Carlo algorithms. 
The quark masses are tuned to their physical values so that, in particular, 
$m_\pi\approx 135 \textmd{ MeV}$ in the vacuum. The error of the pion mass used in 
the simulations originates primarily from the uncertainty of the lattice scale and 
amounts to $2\%$. 
Further details of our lattice action and our simulation algorithm are given in Refs.~\cite{Aoki:2005vt,Borsanyi:2010cj,Endrodi:2014lja}.

Here we perform simulations on a $24^3\times32$ ensemble with lattice
spacing $a\approx 0.29 \textmd{ fm}$, a wide range of chemical potentials 
$0<\mu_I/m_\pi\le 2$ and three pionic source parameters $0.17\le \lambda/m_{ud}\le 0.88$.
The systematic uncertainties originating from lattice artefacts and from neglecting 
$\mathcal{O}(e^2)$ electromagnetic effects will be investigated in a future publication. 
The volume of our system is around $7 \textmd{ fm}^3$, sufficiently large so that 
finite size effects are under control. The temperature is significantly below the relevant QCD
scales so that it well approximates $T=0$.

The isospin density and the pion condensate are obtained as derivatives of the 
partition function~\cite{Brandt:2017oyy},
\begin{align}
\label{eq:defni}
n_I&= \frac{1}{V_4}\frac{\partial \log\Z}{\partial \mu_I}
=\frac{1}{2V_4}\expv{\!\textmd{Re}\,\tr \frac{[\slashed{D}(\mu_I)+m_{ud}]^\dagger \slashed{D}'\!(\mu_I)}{|\slashed{D}(\mu_I)+m_{ud}|^2+\lambda^2}}\,, \\
\sigma_\pi &= \frac{1}{V_4}\frac{\partial \log\Z}{\partial \lambda}
=\frac{\lambda}{2V_4}\expv{\tr \frac{1}{|\slashed{D}(\mu_I)+m_{ud}|^2+\lambda^2}}\,,
\label{eq:defpi}
\end{align}
where the prime denotes differentiation with respect to $\mu_I$. 
Similarly, the axial vector condensate reads
\be
\sigma_A 
=\frac{\lambda}{2V_4}\expv{\tr \frac{U_4}{|\slashed{D}(\mu_I)+m_{ud}|^2+\lambda^2}}\,,
\label{eq:defA}
\ee
where $U_4$ is the staggered equivalent of the timelike component of the 
continuum axial vector operator~\cite{Kilcup:1986dg} that has also been 
used in Ref.~\cite{Bali:2018sey}. In Eqs.~(\ref{eq:defni})--(\ref{eq:defA}), 
$V_4=V/T$ is the four-dimensional volume of the system that includes the 
spatial volume $V=(N_s a)^3$ and the temperature $T=(N_ta)^{-1}$ in terms of the 
lattice spacing $a$ and the lattice geometry $N_s^3\times N_t$. 
Having measured the observables using different values of the pionic source parameter
$\lambda$, the physical results are obtained via an extrapolation to $\lambda=0$.
This is facilitated by using
the singular value representation introduced in Ref.~\cite{Brandt:2017oyy} for 
the pion condensate, which we work out here for $\sigma_A$ as well.

Using the singular values $\xi_n$ of the massive Dirac operator,
\be
|\slashed{D}(\mu_I)+m_{ud}|^2 \psi_n = \xi_n^2 \psi_n\,,
\ee
the pion condensate is rewritten as
\be
\begin{split}
\sigma_\pi&=\frac{\lambda}{2V_4}\Big\langle\sum_n (\xi_n^2+\lambda^2)^{-1}\Big\rangle \\
&\xrightarrow{V\to\infty} \frac{\lambda}{2} \Big\langle\int \!\dd \xi \,\rho(\xi)\, (\xi^2+\lambda^2)^{-1} \Big\rangle \\
&\xrightarrow{\lambda\to0} \frac{\pi}{4}\expv{\rho(0)}\,,
\end{split}
\label{eq:BCpi}
\ee
where we performed the thermodynamic limit introducing the spectral density $\rho(\xi)$, 
followed by the $\lambda\to0$ limit. 
This equation is the analogue of the Banks-Casher relation~\cite{Banks:1979yr}, connecting the order parameter 
of pion condensation to the density of singular values around the origin. 
The determination of $\rho(0)$ involves calculating the low singular values, 
building a histogram and fitting it to extract the spectral density at zero. 
In addition, a leading-order reweighting of the configurations to
$\lambda=0$ is performed. 
For more details on our fitting strategy, see Ref.~\cite{Brandt:2017oyy}. 

A very similar Banks-Casher-type relation can be found for $\sigma_A$ as well. The same 
steps as in Eq.~(\ref{eq:BCpi}) lead in this case to
\be
\begin{split}
\sigma_A &=
\frac{\lambda}{2V_4} \Big\langle\!\sum_n (\xi_n^2+\lambda^2 )^{-1}\psi_n^\dagger U_4 \psi_n  \Big\rangle \\
&\xrightarrow{V\to\infty} \frac{\lambda}{2}\Big\langle\int\!\dd\xi\,\rho(\xi) \, (\xi^2+\lambda^2)^{-1} \psi_n^\dagger U_4 \psi_n \Big\rangle \\
&\xrightarrow{\lambda\to0}
\frac{\pi}{4} \big\langle\rho(0)\,\psi_0^\dagger U_4 \psi_0 \big\rangle.
\end{split}
\ee
The matrix elements of $U_4$ are measured together with the singular values
and extrapolated towards the low end of the spectrum to find $\psi_0^\dagger U_4\psi_0$. 
The so obtained results for $\sigma_A$ are shown in Fig.~\ref{fig:sigma_a} of the body of the text.

We mention that the ratio of the axial vector condensate and the pion condensate 
can also be found from the axial Ward-identity. For $\mu_I\neq0$ but $\lambda=0$, this
operator identity reads
\be
\partial_\nu \bar\psi \gamma_\nu\gamma_5 \tau_a \psi =2 m_{ud} \bar\psi \gamma_5\tau_a \psi +i
\epsilon_{ab3}\,\mu_I \,\bar\psi \gamma_0\gamma_5 \tau_b \psi \,.
\ee
Integrating in space and time, exploiting the periodic boundary conditions for 
the composite field $\bar\psi\psi$ in all directions and taking the expectation 
value over quarks and gluons we get for the $a=2$
component
\be
\mu_I \sigma_A = -m_{ud} \sigma_\pi\,,
\ee
which is found to be satisfied within our statistical errors. (Notice that 
in our definitions $\sigma_\pi$ is related to $i\bar\psi\gamma_5\tau_2\psi$ and 
$\sigma_A$ to $\bar\psi\gamma_0\gamma_5\tau_1\psi/2$.)

\section{Equation of state and the TOV equations}
\label{app:2}

In $\chi$PT the isospin density reads~\cite{Son:2000xc},
\be
n_I = \frac{\mu_I f_\pi^2}{2} \left[ 1-\frac{m_\pi^4}{\mu_I^4}\right]
\cdot\Theta(\mu_I-m_\pi)\,,
\label{eq:nIchiPT}
\ee
where $f_\pi$ is the pion decay constant, which is the only parameter that we allow to vary
for the $\chi$PT fit depicted in Fig.~\ref{fig:nI}.
For free relativistic leptons, the density is
\be
n_\ell(\mu_\ell) = \frac{1}{3\pi^2} (\mu_\ell^2-m_\ell^2)^{3/2} \cdot \Theta(\mu_\ell-m_\ell)\,.
\label{eq:nldef}
\ee
The neutrino density $n_{\nu_\ell}(\mu_{\nu_\ell})$ is analogous to 
Eq.~(\ref{eq:nldef}), just replacing the lepton mass $m_\ell$ by zero and dividing by a factor of 2 since only left-handed neutrinos are considered.
The pionic pressure and energy density are calculated from $n_I(\mu_I)$ at zero temperature via
\be
p=\frac{\log\Z}{V_4} = \int_0^{\mu_I} \!\!\dd \mu_I' \,n_I(\mu_I'), \quad\quad \epsilon= -p+\mu_I n_I\,,
\label{eq:QCDpe}
\ee
and similarly for the charged leptons and the neutrinos, using $n_\ell(\mu_\ell)$ and 
$n_{\nu_\ell}(\mu_{\nu_\ell})$, respectively. 

After requiring local charge neutrality $n_\ell=n_I$, the pion-lepton system is 
unambiguously characterized by the lepton chemical potential $\mu_\ell$. 
The total pressure $p$ and energy density $\epsilon$ enter the TOV 
equations~\cite{Tolman:1939jz,Oppenheimer:1939ne}, which can be rewritten 
in terms of the chemical potentials as
\begin{equation}
\label{eq:TOV_mu}
	\frac{\mathrm{d}\mu_\ell}{\mathrm{d}r} 
	= -G \mu_\ell\, \frac{M + 4\pi r^3 p}{r^2-2rGM}
	\left[1 + \frac{\mu_I}{\mu_\ell} \right]
	\left[1 + \frac{n_\ell'}{n_I'}\right]^{-1},
\end{equation}
where $G$ is Newton's constant, the primes denote derivatives with respect to the corresponding chemical potentials, we used natural units $c=\hbar=1$
and 
\be
	M(r) = 4 \pi \int_{0}^{r} \!\mathrm{d}r'\, r'^2 \epsilon (r'),
	\label{eq:TOV2}
\ee
is the integrated mass. 
Eqs.~(\ref{eq:TOV_mu})-(\ref{eq:TOV2}) remain unchanged if neutrinos are included in the EoS, only $p$ and $\epsilon$ need to be complemented by the respective neutrino contributions.
The first TOV equation (\ref{eq:TOV_mu}) for two lepton families
($\ell=e,\mu$) takes the form
\be
\label{eq:TOV_full}
\begin{aligned}
    \frac{\mathrm{d} \mu_{e}}{\mathrm{d}r}
    &= -G \mu_e \frac{M+4\pi r^3p}{r^2-2rGM}
    \\& \phantom{=}~\cdot
    \left[1 + \frac{\mu_I}{\mu_{e}} + \rho
\left(\frac{\mu_\mu}{\mu_e}-1\right)
    \right]
    \\ & \phantom{=}~\cdot
    \left[1 + \frac{n_{e}'}{n_I'} +
\frac{n_{\mu}'}{n_I'}\frac{\mathrm{d} \mu_{\mu}}{\mathrm{d} \mu_{e}} +
\rho \left(\frac{\mathrm{d} \mu_{\mu}}{\mathrm{d} \mu_{e}}-1\right)
    \right]^{-1},
\end{aligned}
\ee
with asymmetry $\rho = (n_\mu + n_{\nu_\mu})/(n_e + n_{\nu_e} + n_\mu +
n_{\nu_\mu})$ between the lepton families. Note that for $\mu_\mu =
\mu_e$ or $\rho = 0$ the last terms in both square brackets vanish.

The TOV equations are integrated numerically up to the star boundary $r=R$,
where $p_\pi+p_\ell$ vanishes and the total mass $M=M(R)$ is attained. 
Note that for pion-lepton-neutrino configurations, the neutrino pressure is nonzero 
at the so defined boundary.
The points of the mass-radius curves in Fig.~\ref{fig:mr1} correspond to 
different values of the central energy density $\epsilon_c$.

The gravitational stability of the solutions is investigated 
by looking at unstable radial modes. 
In particular, we integrate the Sturm-Liouville equation with oscillation
frequency $\omega = 0$ and check whether the resulting oscillation
amplitude has nodepoints within the star. If so, then there exists at
least one frequency $\omega^2 < 0$, driving the system unstable on long
timescales~\cite{Bardeen}. In the $\pi \ell \nu_\ell$ case, the
integration of the Sturm-Liouville equation was only performed
up to the boundary of the pion condensate. Nevertheless, this was
sufficient to observe whether there are unstable modes (nodepoint within the
pion condensate), or indications of it (no nodepoint, but a clear 
tendency for it within the surrounding neutrino cloud). 
This approach ruled out some configurations that seemed stable
according to the necessary (but not sufficient) condition~\cite{glendenning2000compact},
$\dd M\!(R)/\dd \epsilon_c>0$. \vspace*{.7cm}

\section{Weak decay in the condensed phase}
\label{app:3}

As discussed in Sec.~\ref{sec:ew}, the condensed phase exhibits massive 
photons and no light charged pionic degrees of freedom. Thus, 
standard approaches involving the weak decay of a momentum eigenstate pion do not apply. 
Instead, we need to consider
the production of a lepton pair $\ell(\mathbf{k})\nu_\ell(\mathbf{q})$ out of the condensed ground state $\Omega$, where $\mathbf{k}$ and $\mathbf{q}$ denote the momenta of the charged 
antilepton and of the neutrino,
respectively.\footnote{Remember that we assumed the condensate to carry positive electric charge, 
in which case a charged antilepton and a neutrino are produced in this process.}
As mentioned in Sec.~\ref{sec:ew}, if neutrinos are trapped and the weak 
equilibrium condition $\mu_{\nu_\ell}=\mu_I+\mu_\ell$ is satisfied, the decay at zero temperature
is Pauli-blocked
since all final lepton states are filled. 

For completeness, here we consider the situation, where neutrinos are absent and all final 
lepton states are available. 
According to Fermi's golden rule, the differential probability for the decay process
is related to the $S$-matrix element~\cite{schwartz2014quantum}
\be
\dd P = \sum_{s_\ell,s_{\nu_\ell}}  \frac{|\langle \Omega \ell\nu_\ell|S|\Omega \rangle|^2}{\langle \Omega| \Omega\rangle^2  \langle \ell | \ell\rangle \langle \nu_\ell | \nu_\ell \rangle} \,\left[\frac{V}{(2\pi)^3}\right]^2 \dd^3 \mathbf{k}\, \dd^3 \mathbf{q}  \,,
\label{eq:dPdef}
\ee
involving a sum over the spins $s_\ell$ and $s_{\nu_\ell}$ of the outgoing leptons.
For regularization purposes 
we need to assume that the decay proceeds in a finite volume $V$ and 
over a finite time interval $\mathcal{T}$.

The ground state has unit norm, $\langle \Omega|\Omega\rangle=1$, 
while the normalization of the lepton states takes 
the usual form,
\be
\langle \ell | \ell \rangle = 2E_\ell V, \quad\quad \langle \nu_\ell | \nu_\ell\rangle =2E_{\nu_\ell} V\,,
\ee
where the leptons are on shell,
\be
E_\ell=\sqrt{\mathbf{k}^2+m_\ell^2}, \quad\quad E_{\nu_\ell}=|\mathbf{q}|\,.
\label{eq:onshell}
\ee

The $S$-matrix element factorizes into leptonic and hadronic parts,
\be
\begin{split}
\langle \Omega \ell&\nu_\ell|S|\Omega \rangle
= \frac{G_{\rm F}\cos\vartheta_c}{\sqrt{2}} \\
&\;\cdot \langle \ell \nu_\ell|\bar\ell\gamma^\mu(1-\gamma_5)\nu_\ell|0\rangle \cdot
\langle \Omega | \bar u \gamma_\mu(1-\gamma_5)d|\Omega\rangle\,,
\end{split}
\ee
where $G_{\rm F}$ is the Fermi constant and $\vartheta_c$ the Cabibbo angle. 
The leptonic component can be treated as usual~\cite{okun2013leptons}.
The hadronic factor reflects the accumulation of weak vertices in the pion condensate. 
While the expectation value of the vector part vanishes, $\expv{\Omega|\bar u \gamma_\mu d|\Omega}=0$, the zeroth 
component of the axial vector part is nonzero. In $\chi$PT it reads\footnote{In fact, $\sigma_A$ and $\sigma_\pi$ 
are orthogonal to each other in isospin space, so that the axial vector current 
is parallel to the would-be Goldstone mode, see also Ref.~\cite{Brauner:2016lkh}. Here we chose the direction of spontaneous 
symmetry breaking such that $\sigma_\pi=\langle\bar u \gamma_5 d - \bar d \gamma_5 u\rangle$,
implying $\sigma_A=\langle\bar u \gamma_0\gamma_5 d+\bar d \gamma_0 \gamma_5 u\rangle/2$. 
Note also that for each observable, the two contributing terms are equal in magnitude.}~\cite{Brauner:2016lkh}
\be
\langle \Omega| \bar u \gamma_0 \gamma_5 d|\Omega\rangle
= \sigma_A =  -\frac{m_\pi^2f_\pi^2}{2\mu_I} \sqrt{1-\frac{m_\pi^4}{\mu_I^4}}
\cdot\Theta(\mu_I-m_\pi)\,,
\ee
which is plotted in Fig.~\ref{fig:sigma_a} together with the corresponding lattice data.

Energy conservation implies $\mu_I=E_\ell+E_{\nu_\ell}$, since after the decay the charge of 
the condensate is reduced by one unit, releasing $\mu_I$ energy. 
To maintain the zero-momentum state of $\Omega$, we assume that the condensate picks up zero momentum so that momentum conservation fixes
$\mathbf{k}=-\mathbf{q}$.
Performing the spin sums in the leptonic factor, 
the squared matrix element for $\mu_I>m_\pi$ becomes 
\be
\begin{split}
\hspace*{-.3cm}\sum_{s_\ell,s_{\nu_\ell}}\! |\langle \Omega \ell&\nu_\ell|S|\Omega \rangle|^2
\!= 4 \,G_{\rm F}^2\cos^2\!\vartheta_c \,\sigma_A^2 \,(E_\ell E_{\nu_\ell} +\mathbf{k}\cdot\mathbf{q}) \\
&\cdot (2\pi)^4\delta(\mu_I-E_\ell-E_{\nu_\ell})\,\delta^{(3)}(\mathbf{k}+\mathbf{q}) \,V\mathcal{T}\,,
\end{split}
\ee
where the regularization $(2\pi)^4\,\delta(0)\,\delta^{(3)}(\mathbf{0})=V\mathcal{T}$ was used.

Inserting this in Eq.~(\ref{eq:dPdef}), performing the integrals over
the momenta and using the on-shell conditions~(\ref{eq:onshell}), the decay rate $\Gamma=\int\! \dd P/\,\mathcal{T}$ 
in the condensed phase $\mu_I>m_\pi$ reads
\be
\Gamma = \Gamma_{\rm vac} \cdot \frac{m_\pi^3}{\mu_I^3} \left[\frac{1-m_\ell^2/\mu_I^2}{1-m_\ell^2/m_\pi^2}\right]^2 \!\cdot n_Q    V\,,
\label{eq:Gammares}
\ee
where $\Gamma_{\rm vac}$ is the decay rate of a pion at rest in the vacuum 
(see, e.g., Ref.~\cite{okun2013leptons})
and we factored out the density $n_Q=n_I$ using Eq.~\eqref{eq:nIchiPT}. 
Thus, for high isospin chemical potentials, $\Gamma$ is reduced as $\mu_I^{-3}$.
The result~(\ref{eq:Gammares}) is extensive in the volume, since the weak current can couple to the 
condensate at any point in space. Keeping the number 
of charges $N_Q=n_QV$ fixed, in the limit $\mu_I\to m_\pi$ the decay rate 
reproduces $N_Q$ times the vacuum decay rate, satisfying continuity of $\Gamma$ at the pion 
condensation onset. Altogether, the average lifetime of the condensate 
thus reads $\tau = (\Gamma/N_Q)^{-1}$, giving Eq.~(\ref{eq:taures}) in the 
body of the text. 

\bibliographystyle{utphys}
\bibliography{pistar}

\end{document}